\documentclass[conference]{IEEEtran}
\pdfoutput=1
\usepackage[font=bf]{caption}

\usepackage{times}
\usepackage{xspace} \usepackage{epsfig}
\usepackage{amsmath}
\usepackage{amssymb}
\usepackage[hyphens]{url}
\usepackage{listings}
\usepackage{fancyvrb}
\usepackage{graphicx}
\usepackage{epstopdf}
\usepackage{url}
\usepackage{multirow}
\usepackage{float}
\usepackage{subfig}
\newcommand{\subparagraph}{}
\usepackage{algorithm,algorithmicx,algpseudocode}
\usepackage{varwidth}
\usepackage[compact]{titlesec}
\usepackage[usenames,dvipsnames]{color}
\usepackage{mathtools}
\usepackage{amssymb}
\usepackage{pifont}
\definecolor{purple}{rgb}{0.59, 0.44, 0.84}
\newcommand{\comment}[1]{}

\newcommand{\mypara}[1]{\smallskip\noindent{\bf {#1}:}~}

\newcommand{\ignore}[1]{}

\newcounter{packednmbr}

\newenvironment{packedenumerate}{\begin{list}{\thepackednmbr.}{\usecounter{packednmbr}\setlength{\itemsep}{0.5pt}\addtolength{\labelwidth}{-4pt}\setlength{\leftmargin}{\labelwidth}\setlength{\listparindent}{\parindent}\setlength{\parsep}{1pt}\setlength{\topsep}{0pt}}}{\end{list}}

\ifCLASSINFOpdf
\else
\fi
\begin{document}
%
\title{Shedding Light on the Adoption of Let's Encrypt}




%
\author{\IEEEauthorblockN{Antonis Manousis,
Roy Ragsdale,
Ben Draffin,
Adwiteeya Agrawal and 
Vyas Sekar}
\IEEEauthorblockA{School of Electrical and Computer Engineering\\
Carnegie Mellon University\\ 
\{antonis,  bendraffin, adwiteeya, rragsdale,\}@cmu.edu}}



\maketitle
\begin{abstract}

Let's Encrypt is a new entrant in the Certificate Authority ecosystem that
offers free and automated certificate signing. It is visionary in its commitment
to Certificate Transparency. 
 In this paper, we shed light on the adoption  patterns of 
 Let's Encrypt ``in the wild'' and inform the future design and deployment
 of this exciting   development in the security landscape.  
 We analyze acquisition patterns of certificates as well as their 
 usage and deployment trends in the real world. 
 To this end, we analyze data from
Certificate Transparency Logs containing records of more then 18 million
certificates. We also leverage other sources like Censys, Alexa's historic records,
Geolocation databases, and VirusTotal. We also perform active HTTPS measurements on the
domains owning Let's Encrypt certificates.
 Our analysis of certificate acquisition  shows that (1) the impact of Let's Encrypt is particularly visible
in Western Europe; (2) Let's Encrypt has
the potential to democratize HTTPS adoption  in countries that are recent
entrants to Internet adoption; (3) there is anecdotal evidence of  popular
domains quitting their previously untrustworthy or expensive CAs in order to
transition to Let's Encrypt; and (4) there is a ``heavy tailed'' behavior 
 where a small number of domains acquire a large number of certificates.
 With respect to usage, we find that: (1)  only 54\% of domains
actually use the Let's Encrypt certificates they have procured;  (2)
there are many non-trivial
incidents of server misconfigurations; and (3) there is early evidence of
use of Let's Encrypt certificates for typosquatting and for
malware-laden sites. 
Based on these results, we derive key security implications and recommendations 
 for Let's Encrypt, website administrators, browser vendors, and end users. 

\end{abstract}

\section{Introduction}

Today's web ecosystem critically relies on  Certificate Authorities (CAs) in order to
ensure that network connections are trusted and secure, as they  vouch for the
binding between a domain name and its public key by issuing digital
cryptographic certificates.  While CAs are a key part of this ecosystem, there are 
 several known problems with them.  First, the issuance of SSL/TLS certificates by
traditional Certificate Authorities has historically been a manual and
expensive process, with costs (per certificate) somewhere between \$$5$ and
\$$1000$ a year~\cite{cheap, expensive}. Second, Certificate Authorities have had little
to no incentive for transparency so they rarely publish complete lists of the
certificates they have signed; instead they serve as the sole entity to
evaluate certificate requests. This has serious security ramifications in
the event a compromised or malicious CAs issues fraudulent and untracked
certificates. This threat is not hypothetical and has precedent: DigiNotar  mis-issued SSL certificates that were later used to
perform man-in-the-middle attacks against web users~\cite{diginotar}. 

Certificate Transparency is a new effort that promises to combat the danger of
compromised or malicious CAs issuing rogue certificates by adding all
newly issued or revoked certificates to a public, verifiable, append-only
log~\cite{google}.  This ensures that if a certificate is not in the log, the
client can take appropriate action and refuse to connect to a particular
website. If certificates are in the log, the community can quickly
determine if they are dangerous or fraudulent and act to revoke them. 
The development of Certificate Transparency has contributed to the
emergence of (i) protocols that automatically sign minimally-validated
certificates such as the Automated Certificate Management Environment
(ACME)~\cite{ACME} and (ii) Certificate Authorities that utilize these
protocols to automate the procedure of issuing certificates, like Let's
Encrypt~\cite{LE}.  Let's Encrypt is currently the primary Certificate Authority
leveraging the ACME protocol. Let's Encrypt appears to be particularly popular as it is currently
signing certificates at a rate of about 55,000 a day.  A particularly
interesting feature   of Let's Encrypt is that it is both free and automated.

\mypara{Motivating Questions} The goal of this paper is to shed light on the
{\em adoption} patterns of Let's Encrypt certificates ``in the
wild''.  Such an understanding  can be immensely useful to multiple players in
the end-to-end HTTPS ecosystem. For instance, this can shed light on: (1) how
and where  HTTPS adoption can be further stimulated; (2) expose potential
sources of design optimizations and considerations for future deployments
(e.g., are there flash crowds or heavy tailed behaviors); and (3) understand
if/how the free and automated deployment can lead to potential sources of
misconfigurations or abuse.

We divide our analysis into two high-level categories: (1) {\em
  acquisition} of certificates and (2) {\em usage} in the~wild.
\begin{enumerate}

\item {\em Acquisition:} With respect to  acqusition, we seek to analyze several natural questions 
 with respect to:
\begin{itemize}

\item  Geographic characteristics of issued certificates: \\
Is Let's Encrypt 
 popular in countries with already high web/HTTPS penetration or is it more popular 
 in countries with emerging Internet adoption? 

\item The `profile' and motives of web domains that obtain them: \\
 Is adoption 
 uniformly popular across different users or are some acquiring more 
 certificates than others?  Were sites previously using HTTPS with
 other CAs and then switching to Let's Encrypt? How many of the certificates
 are first time HTTPS deployments? Are the early adopters popular websites or in the ``tail'' of the popularity 
distribution?

\end{itemize}

\item {\em Usage:} With respect to usage patterns in the wild, there are  several natural questions 
 regarding: 
\begin{itemize}

\item Actve deployment: \\ 
 Are  users requesting Let's Encrypt certificates 
merely out of curiosity or are they being deployed in ``production''?  Are users using these certificates 
 and renewal processes correctly?


\item Malicious Intents: \\
Does the free and automatic character of issuing trusted certificates become an enabler 
 for malicious activites such as typosquatting and for delivering malware?

\end{itemize}
\end{enumerate}


 

\mypara{Methodology and Findings}
Using a dataset of around 18 million certificate logs as well as Alexa's 
records of the top 1 million web domains and services like Censys, VirusTotal and Geolocation
databases we set out to answer questions on adoption and usage. We use the Certificate Transparency
Logs in order to identify the domains certified by Let's Encrypt and then we collect amplifying 
information from the aforementioned services. 

Using this data, we  perform our measurement
experiments and our key findings are as follows:

\begin{itemize}
\item \textbf{Geographical distribution of certificates:} 
Let's Encrypt has notably contributed to the democratization of adoption 
of TLS certificates. In countries like Argentina, Ukraine and South 
Africa, Let's Encrypt certificates are over-represented in the whole population of 
HTTPS websites by a factor of 5. In western Europe, countries like Switzerland, 
France and the Netherlands issue the most certificates as a function of their
active Internet population with an average ratio of 3.9 certificates per thousand
active Internet users.    
\item \textbf{Characteristics of adopting websites:}
We investigate the characteristics of websites that have adopted Let's Encrypt as
their CA. We observe that 1.2\% of popular websites appearing in Alexa top million
have transitioned from paid CAs to Let's Encrypt. 

\item \textbf{Bulk acquisition of certificates:} We estimate that around 7\% of all issued
Let's Encrypt certificates are used by companies for services that provide web-interfaces for 
secure communication between end-users and the company's products (e.g. routers, Network Attached 
Storage Devices or dynamic DNS servers). We identify and discuss implications of this practice. 

\item \textbf{Usage characteristics:} 
The free nature of Let's Encrypt has inadvertently created a pool of users that are 
interested in using the platform only out of curiosity. We estimated that the number of 
domains for which multiple redundant certificates have been issued is on average
9\% of the number of domains using Let's Encrypt. Further, we found that 15\% of the total number 
of Let's Encrypt domains opt out by not renewing their certificates. Finally, we observe 
that of the active domains that obtained a Let's Encrypt certificate,
only 50\% replied with a valid Let's Encrypt certificate on the
standard HTTPS port.
\item \textbf{Malicious usage: } 
We see anecdotal cases where users have tried to provide legitimacy to malicious websites by
`securing' them with certificates provided by Let's Encrypt. We experiment with typosquatted
domains and discover that indeed malicious users have tried to leverage free TLS certificates
to exploit users' trust.
\end{itemize}

\mypara{Implications} Based on the  findings, we derive some  key implications and recommendations
 for Let's Encrypt,  domain owners, users,    and for browser vendors:
\begin{enumerate}
\item {\em For Let's Encrypt:}  First, the adoption patterns of Let's Encrypt suggest that there is a number of domains requesting a 
 large number of certificates for subdomains. This could both increase the load for renewals on Let's Encrypt 
 and needlessly bloat the audit logs. Therefore, Let's Encrypt should maybe revisit the decision to not support
 wildcard certificates (e.g., *foo.com), for cases where they would be a secure option (and provide
guidance to domain owners on when wildcard certificates are not a suitable option). Second, the early signs 
 of free certificates being used for malicious intent suggest that Let's Encrypt can run  checks for malicious indicators 
 or for typosquatting intent  as a  proactive measure before issuing certificates. 

\item {\em For domain owners:}  Domain owners could benefit from some simple sanity checks to avoid 
 common misconfiguration errors as we observed above. We also suggest that domain owners refrain 
 from requesting certificates for inactive domains to avoid bloating up the audit logs. 
 Finally, domain owners could benefit the community by providing more metadata and being more forthcoming 
 in giving reasons or their intents while requesting certificates; e.g., why domains are inactive or why they transitioned.

\item {\em For users and browser vendors:} Given that there are a few common misconfiguration templates (e.g., default 
 ``toy'' server configuration),   users and browsers  could
 proactively check for these when accessing HTTPS-enabled sites and
 potentially report them to a central ``certificate health'' repository.

\end{enumerate}

\mypara{Roadmap} Section 2 provides some background on Certificate
Transparency, Let's Encrypt and related measurement studies. Section 3 discusses the dataset and methodology used in our measurements.
Sections 4 and 5 analyze the results of the adoption and usage patterns
respectively. Section 6 presents implications of our analysis and finally
Section 7 concludes this work.
\section{Background and Related Work}
\label{sec:back}
In this section we first provide background on Certificate Transparency and 
the mechanisms of Let's Encrypt. Then, we discuss previous works that have 
provided preliminary measurements and insights on the adoption of Let's Encrypt
 
\subsection{Certificate Transparency}

The core idea behind Certificate Transparency is to create a public,
verifiable, append-only certificate log. The log's integrity can be trusted
because all of its entries are cryptographically verifiable. By checking the
log, any client can verify the validity of a certificate before connecting to a
particular website. If the certificate is not logged, then the client can reject
the connection and avoid mistrusting the provided certificate. Appending
certificates to the log does not add any extra latency as log-inclusion proofs
are incorporated to the TLS handshake~\cite{google}. 

To avoid Certificate Transparency 
logs being wasteful in storage and bandwidth, current approaches leverage Merkle Trees.
These are efficient data structures in which a leaf is an item in the
log and each branch is a cryptographic hash of the 
nodes below the branch. The root of the tree, assembled by successive
hashing, is therefore a summary of all it's contents. 
A site may provide the signed root node of the transparancy log's Merkle Tree
to prove that it has not changed the log or misbehaved~\cite{google}. 
This allows for trusted and efficient log-inclusion proofs. The 
proliferation of an HTTPS ecosystem 
in which Certificate Transparency is an integral component 
makes scalability considerations important factors in the design of
logs. As the web scales, they need to keep vast 
amounts of data and also to synchronize them between different replicas worldwide.

\subsection{Let's Encrypt}

The principal goal of Let's Encrypt~\cite{LE} and the ACME protocol~\cite{ACME,
ACME_git} is to make security accessible for everyone. The strategy
for doing so is a service for websites to obtain a browser-trusted certificate without any
human intervention by the Certificate Authority and minimal effort
from the server's administrator. 
This happens by running a certificate management agent on the web
server.  The procedure happens in two steps. First, the agent proves
to the CA that the web server controls a domain and then the agent can
request, renew, and revoke certificates for that domain.

Let's Encrypt uses public key cryptography to identify the server. During the
first communication between the server and Let's Encrypt, a new public-private
key pair is generated. This is similar to the process of creating an account at
a traditional CA. The server administrator queries Let's Encrypt in order to
find out how the user should prove they own the domain. The CA then issues
one or more challenges. The current options are; (1) check DNS records for the
domain or (2) access a token on a specific URI on that domain. Additionally,
Let's Encrypt issues a nonce that the agent running on the web-server needs to
sign with their newly-issued private key~\cite{LE}. Once the challenges have been 
satisfied, the domain has been validated, and the key pairs are authorized. The 
user can now issue requests for new certificates as well as renew or revoke existing 
certificates. They must simply send certificate management messages signed with 
the authorized key pair. 

One interesting point about certificate issuance is choosing the log
that will keep a record of the certificate. While Let's Encrypt claims to submit
all its certificates to Certificate Transparency Logs, there is no
stated official log that
they submit to~\cite{lelog}. Information about published certificates can be found 
on domains specializing in certificate searches like~{\tt https://crt.sh}.  

\subsection{Related work}

\mypara{Early measurements of Let's Encrypt}
A parallel effort to quantify the impact of Let's Encrypt has been performed by
J.C. Jones, a Security Engineer at Mozilla who designed the infrastructure for
Let's Encrypt and currently serves on its Technical Advisory Board
~\cite{124days, articl}. This work utilized Certificate Transparency Logs as the
primary lens to gain insight into the adoption and utilization of Let's Encrypt
as a Certificate Authority.  It investigated the number of domains using Let's
Encrypt certificates that did not previously have a TLS certificate.  Moreover,
Jones addressed the utility of Certificate Transparency Logs as
a data source by also examining the number of certificates that can be found in
scans by Censys~\cite{censys-ccs15} but not in Certificate Transparency Logs.  Using
this data, the post observes how Let's Encrypt compares with other established
Certificate Authorities, such as Comodo, Symantec or GoDaddy.

This concurrent work determined that $90.4\%$ of the domains using Let's Encrypt
are new to Web PKI and Let's Encrypt ranks fourth in the list of most observed
unexpired certificates grouped by issuer.  Also, it presented data showing that
$55\%$ of issued Let's Encrypt Certificates are only found in Certificate
Transparency Logs, $35\%$ can be found in both Transparency Logs and Censys and
$9\%$ can only be found in Censys thus validating the largely comprehensive nature of
Certificate Transparency Logs. 
Interestingly, the presence of sites with Let's Encrypt certificates
not listed in transparancy logs seems counter to the goal that all Let's
Encrypt certificates are submitted to logs. Remedying this disconnect
will be an important issue going forward.

While this parallel effort tackled high level questions concerning the scale of
Let's Encrypt's operations, and demonstrated how a commitment to Certificate
Transparency makes these logs an invaluable lens into an otherwise opaque
ecosystem, it did not attempt to link certificate issuance to actual
usage.  We observe that the unprecedented low barriers to entry
in the automated and free nature of Let's Encrypt causes this simplifying
assumption to not hold. It is invalid to assume that just because
a certificate was issued that it is actively, or correctly, securing a domain.
In this work we start from the same foundation of Certificate Transparency Logs
but go further by leveraging other sources such as Geolocation
databases, Alexa's historical records, VirusTotal, and active HTTPS scans to shed light on
specifically measurable aspects of the adoption and utilization of Let's Encrypt.

\mypara{Related measurement efforts} Our work follows in the line of a
rich history of measurements of security-related aspects of the web ecosystem.
There are several orthogonal efforts analyzing other aspects of the HTTPS
ecosystem.  Liu et~al., study the frequency and accuracy of certificate
revocation and checking; they paint a less-than-optimistic picture of the
effectiveness of revocation~\cite{liu-imc15}.  Durumeric et~al., analyze the
prevalence of the Heartbleed vulnerability in web servers and assess the impact
on the web certificate ecosystem~\cite{durumeric-imc14}.  Zhang et~al, more
specifically look at the issue of revocation behavior in the aftermath of
Heartbleed~\cite{zhang-imc14}.  Bates et~al.,  analyze the adoption of the
Convergence CA extension to the ``crowdsourced'' approach for certificate
verification named Perspective~\cite{wendlandt-usenix08}; they measure 
how effective it will be at
scale and suggest that simple caching strategies can improve
scalability~\cite{bates-imc14}.  Naylor et~al, have analyzed the costs of
websites moving to HTTPs~\cite{naylor-conext14}  and Varvello et~al, analyze
the adoption of the HTTP/2 standard~\cite{varvello-pam16}. In terms of
measurement infrastructures, Zmap~\cite{zmap-usenix14} and
Censys~\cite{censys-ccs15} provide novel capabilities for Internet-scale
scanning and provide useful datasets that enable the kinds of analysis we
perform in this paper. A recent work by Lever~et~al., looks at the notion of residual
 trust when domain registrations expire and are repurposed with a different use~\cite{lever-oakland16}. 
 We observe that a small but non-trivial fraction of certificates were not renewed. 
  Given the automated nature of certificate renewals, an interesting direction for 
 future analysis is measuring scenarios if these certificate renewals failed because 
 domain registrations lapsed.

\mypara{Other related work in web certificates}
Let's Encrypt leverages new ideas like Certificate Transparency and the ACME
protocol. It promises to revolutionize the tedious, human-in-the-loop procedure
of issuing trusted certificates. It is the first organization to automate such
a procedure, yet will likely not be the last. The following subsection focuses on
recent articles that showcase the early  impacts of Let's Encrypt~\cite{articl}. 
For the interested reader, related research in the broader area of certificates 
and web-trust includes but is not limited to works that: evaluate the Certificate Trust
Model~\cite{trust}, discuss efficient `gossip' protocols to ensure consistency
of certificate logs~\cite{gossip},  develop methods for enhanced certificate
transparency and end-to-end encrypted mail~\cite{ryan}, and discuss common SSL
errors or cases of forged certificates~\cite{errors,forged}.

\section{Dataset and Methodology}

In this section we describe our datasets, the methodology by which we collected 
and processed them, as well as the limitations of our approach. Our decisions 
were guided by our overarching goal which is to evaluate the adoption and usage
patterns of Let's Encrypt certificates. Specifically, our
methodology should allow us to investigate the rate of adoption of Let's Encrypt 
certificates, the motives of adopting users, their geographic distribution as well
as the HTTPS history of Let's Encrypt-certified domains. We also needed
be able to identify patterns of misuse or abuse on behalf of users as well as indications
for malicious activities.

\subsection{Sources of Data}

Our first goal was to collect a complete list of domains that have used 
Let's Encrypt as their Certificate Authority. The source of these domains 
was Certificate Transparency logs, which collectivelly contain a
complete record of virtually all the signed certificates issued by Certificate Authorities
complying with Certificate Transparency rules, including Let's Encrypt. 
We evaluated two primary options 
for obtaining certificate transparency logs; a full log mirror solution and a
simpler log downloader.  As a major supporter of Certificate Transparency,
Google provides an open source log server solution that can also be used in
mirror mode~\cite{googlemirror}.  We first examined this option as it would be
the most full-featured solution.  However, community members have
built more streamlined tools for batch downloads, leading us to
a second method, a simple log downloader. We utilized the code 
provided by James `J.C.' Jones~\cite{jcjones} as well as the original ct-sync 
tool from Adam Langley as described in his efforts to establish Certificate 
Transparency~\cite{CTblog}. 

Using this method we downloaded the full set of Certificate 
Transparency Logs from the Certly Log Server~\cite{certly} which includes  
certificates issued by Let's Encrypt up to March 31st. In this batch, we obtained 
certificate logs for 1,331,781 unique certificates to analyze.  This includes 
1,156,266 certificates issued by Let's Encrypt. This full mirror was the first 
dataset which we used for analysis and further amplification. On April 15th 2016, 
Google announced that Certly failed to comply with Certificate Transparency 
regulations, specifically the requirement about 99\% uptime~\cite{group} and 
therefore was no longer considered as a trusted CT log by the community.
We proceeded by scaling our measurements up by downloading 
Google's most recent Aviator CT log of around 17 million entries. As mentioned 
in Section \ref{sec:back} there is no offical Certificate Transparency
log for Let's Encrypt certificates. A comparison between the official
announcements from the Let's Encrypt organization~\cite{LEstats}
about issued certificates with the number of Let's Encrypt certificates appearing in Certly
showed that Certly contained 90\% of the issued certificates up until March 30th and therefore
at the time we evaluated that this would be an adequate preliminary data set. After Certly was no longer
considered a trusted Transparency log, the Aviator log was determined
to be a satisfactory replacement.

The information stored in Certificate Transparency logs was the starting point for 
most of our experiments. This allowed us to bootstrap our 
measurements related with the geographical characteristics of Let's Encrypt, 
typosquatting analyses and also enabled the collection of auxiliary information from 
services like VirusTotal~\cite{VT} and Alexa~\cite{alexa}. The validity period of each certificate 
allowed us to investigate certificate usage 
patterns and user behavior, including determining the existence of duplicate certificates for a particular 
domain. A caveat of this method, however, is that since many traditional Certificate Authorities 
do not make their certificate records public, the knowledge about HTTPS history of domains 
would have to come from other sources.

To collect additional information about each domain found in the Certificate Transparency
logs, we had to leverage other services that could provide us with useful amplifying 
information and metadata. This information came from the following datasets and
allowed us to gain clearer insights into the Let's Encrypt ecosystem.
  
\begin{packedenumerate}
\item \textbf{VirusTotal~\cite{VT}:} an online service which can be used to scan 
a URL, domain name, IP or a file against various  `malicious activity 
detection' services. Specifically, VirusTotal scans each URL with 67
different antivirus solutions and reports how many reported that a domain
is suspicious or infected.
\item \textbf{Censys~\cite{censys}:}  provides internet-wide scan data capturing how devices,
websites, and certificates are configured and deployed. It contains a collection of 
historic data for most domains of interest and provides a SQL like interface for users
to make queries to these sources of data.  
\item \textbf{Alexa~\cite{alexa}:} a service that provides commercial web traffic data and 
analytics in order to benchmark and compare web domains.
\item \textbf{Geolite2 Geolocation Database~\cite{geolite}:} This service allowed us to get geographic information
about domains of interest.
\end{packedenumerate}

\section{Analysis of certificate acquisition }

In this section we present results pertaining to the acquisition patterns of Let's Encrypt 
certificates. First, we evaluate the rate of acquisition of Let's Encrypt and then we proceed
by characterizing the geographic distribution of certificates. We then explore the prevalence of 
certificates among popular websites and finally, we investigate the `profiles' and motives of 
adopting users, especially those who obtain certificates in large quantities.

\subsection{Rate of Adoption}

\begin{figure}[t]
  \centering
    \includegraphics[width=200pt]{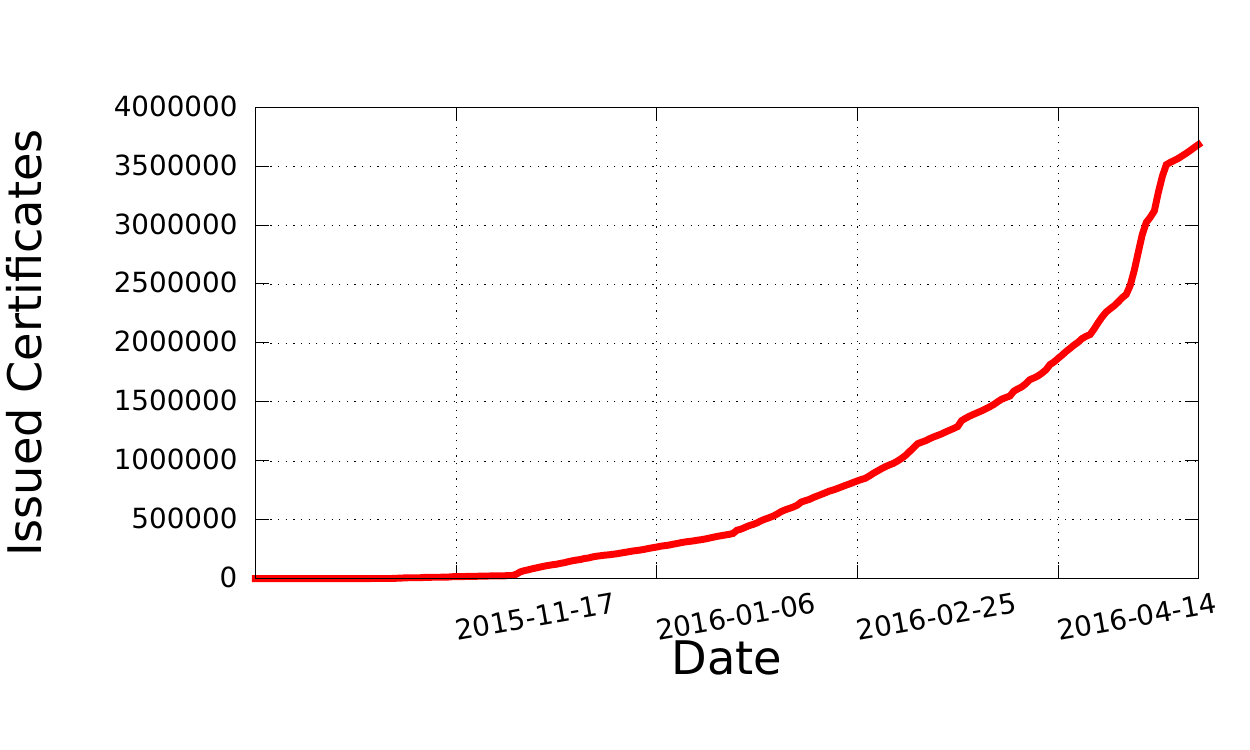}
    \caption{Overall acquisition patterns of LE certificates}\label{fig:adop}
\end{figure}

To analyze the acquisition of  Let's Encrypt as a Certificate Authority, 
we downloaded Certificate Transparency logs from 17 September 2015 up to 15 May 2016. 
Figure~\ref{fig:adop} confirms the popular perception that adoption 
 is on the rise and shows the total number of certificates issued by Let's
Encrypt over time. We can clearly see that acquisition is on the rise with close
to 4M certificates issued  as of May 14, 2016. 

\begin{figure}[h]
\subfloat[Number of certificates]
{
  \centering
    \includegraphics[width=200pt]{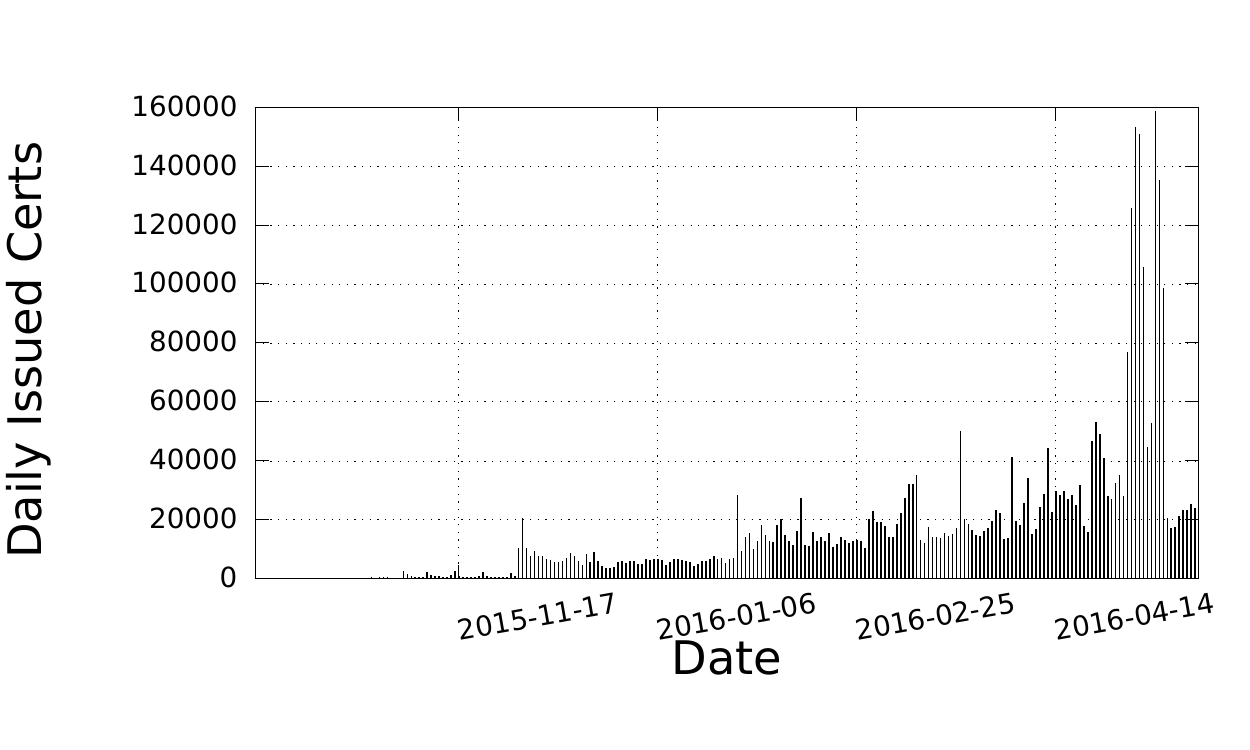}
\label{fig:daily_adopt:cert}
} \\
\subfloat[Number of effective TLDs]
{
    \includegraphics[width=200pt]{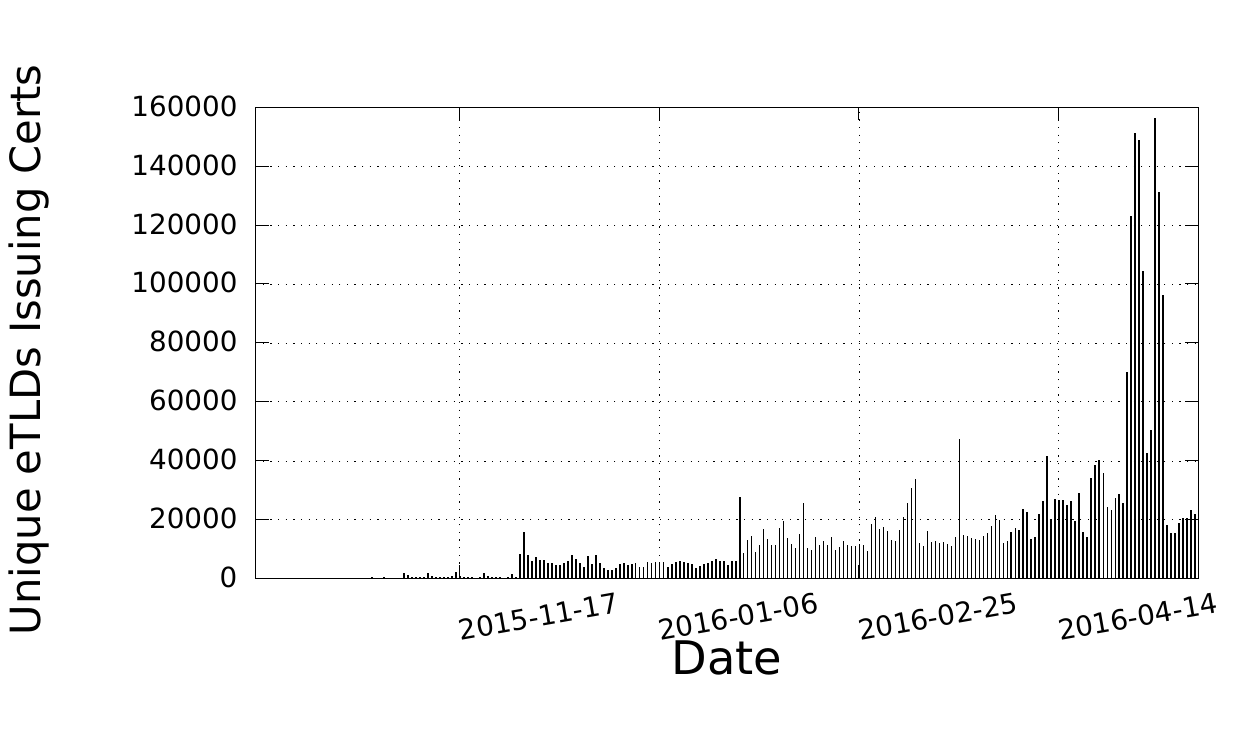}
\label{fig:domain_adopt:dom}
}
\caption{Per-day analysis of certificates and unique Domains requesting certificates}
\label{fig:daily}
\end{figure}

 We also analyzed if there were interesting temporal patterns in the issuance
of these certificates.  To this end, we looked at the number of certificates
issued per day as well as the number of unique effective TLDs (e.g., {\tt xyz.com}
or {\tt foo.co.uk} rather than {\tt .com} or {\tt .co.uk})   that acquired certificates on a
daily basis in Figure~\ref{fig:daily}.  Figure~\ref{fig:daily_adopt:cert} shows
the aggregate number of issued certificates per day, while   Figure~\ref{fig:domain_adopt:dom} shows the unique effective TLDs (eTLDs)
requesting Let's Encrypt certificates on a daily basis.
 This result is generally consistent  with our general acquisition graph but it also reveals an interesting observation that
there are certain abnormal peaks in early May.

We further investigated the root causes of these spikes and found two 
types of explanations. The first observation was that a set of providers
were issuing certificates for many of the subdomains 
they controlled. That is either because Let's Encrypt does not support wildcards;
e.g., {\tt xyz.com} requests certificates 
for {\tt a.xyz.com} and {\tt b.xyz.com} or because each subdomain represents a different user account requiring its own
dedicated certificate. These providers 
generally obtain a very large number of certificates at once. An example of
one such user is {\tt automattic.com}, a web development company that
between May 4th and May 7th 2016 issued 850 certificate for its clients.
Table~\ref{table:highiss} depicts Let's Encrypt users that issued the
highest number of certificates for the peaks of figure~\ref{fig:daily_adopt:cert}. 
Interestingly, the same user is responsible for acquiring certificates
in bulk on a regular basis. 

\begin{table}[h]
\small
\begin{center}
 \begin{tabular}{||c | c | c||} 
 \hline
 Date & eTLD & Issued certificates  \\ [0.5ex] 
 \hline\hline
 May 4th & freeboxos & $5419$ \\
 \hline
 May 5th & freeboxos & $2159$ \\
 \hline 
 May 6th & freeboxos & $1652$  \\
 \hline
 May 7th & freeboxos  & $1426$ \\ 
 \hline
 May 10th& freeboxos  & $1199$   \\
 \hline
\end{tabular}
\caption {eTLD with most certificates issued}\label{table:highiss}
\end{center}
\end{table}

 However, that does not tell the full story and cannot fully explain the spikes. In fact, if we look closely 
 at  Figures~\ref{fig:daily_adopt:cert} and~\ref{fig:domain_adopt:dom},
 what we actually see is that the number 
of unique domains issuing certificates is comparable to the numbers of certificates issued per day. 
 Using DNS resolution and whois lookups on  the set of 
   domains that were issued certificates on these specific days revealed that many of them 
 in fact had the same authoritative name server or the same NS entry in the WHOIS record. 
  A specific example was a French provider {\tt ovh.com} that seemed to have obtained certificates
 for many unique eTLDs that do not share a common suffix; i.e., these were domains of 
 the form {\tt xyz.com} rather than {\tt xyz.ovh.com}. In particular, during the days with the highest
spikes on adoption, we observed that {\tt OVH} was responsible for 30-35\% of the total number of certificates
issued and that up to the end of our measurement period, it had acquired around 650,000 Let's Encrypt certificates.


\subsection{Analysis of Top Users}

The insight that particular users consistently acquire large numbers of certificates
led us to proceed with our investigation of the websites using Let's Encrypt 
clustered by the effective TLD (eTLD). For example,  domains {\tt a.example.com} and {\tt b.example.com} were 
clustered under {\tt example.com}. Similarly, domains such as { \tt a.example.co.uk} and {\tt b.example.co.uk} were clustered under
{\tt example.co.uk}. The goal was to pinpoint cases of domains with a surprisingly
high number of issued certificates.   We excluded cases of domains where the Second Level Domain represents a geographic 
location (for example domains in Ukraine are clustered at a city level so a domain in Kiev will end in {\tt kiev.ua}),
as we wanted to identify unique entities obtaining Let's Encrypt certificates 
in bulk.


\begin{table}[H]
\small
\begin{center}
 \begin{tabular}{||c | c||} 
 \hline
 SLD & \% Issued Certificates   \\ [0.5ex] 
 \hline\hline
 freeboxos & $155,071$ \\
 \hline
 synology & $6109$ \\
 \hline 
 cdn77 & $5928$  \\
 \hline
 hoffman-andrews  & $3114$ \\ 
 \hline
 duckdns  & $2820$   \\
 \hline
 dyndns & $1653$   \\
 \hline
\end{tabular}
\caption {Second Level Domains with most Let's Encrypt certificates}\label{table:SLD}
\end{center}
\end{table}

Table~\ref{table:SLD} contains the 6 highest e-TLDs and the number of certificates issued.
 For the specific entries in Table~\ref{table:SLD},  what we observe anecdotally is that companies often obtain 
 large numbers of Let's Encrypt certificates in order to provide secure connections between the end-user and the web interfaces
of services they sell. For example, `FreeboxOS' is a service offered by the popular low-cost French ISP `Free' that combines
telephony, satellite TV and WiFi~\cite{Free}. FreeboxOS provides a web interface so that users can connect to their local modem and 
safely manage their account, therefore they issue a TLS certificate per userID. Similar patterns are observed by other 
companies in the table above like `duckDNS' and `Synology'. (The one exception is  Jacob Hoffman-Andrews who 
 is a programmer, tech blogger  and contributor to Let's Encrypt which explains the high number of Let's Encrypt certificates issued for his personal website.)

The certificates acquired by the domains in Table~\ref{table:SLD} account for almost 7\% of the total number of certificates 
issued by Let's Encrypt. Their purpose is to provide secure access to individual client-owned devices  
(e.g freeboxos cable boxes), as well as to user profiles managed by the certificate-acquiring organizations.
The service for which these certificates are issued requires individual, per-user certificates, rather than wildcard
certificates (*.domain.com) which would make users susceptible to Man-In-The-Middle attacks. However, the observation that 
companies request very high numbers of certificates  also draws attention to the fact that Let's Encrypt does 
not currently support wildcard certificates. 

Supporting wildcard certificates is, in our opinion, a feature that would be useful for specific applications.
For example, libraries around the world use a product called EZproxy to connect library patrons to licensed resources. In order 
for the connection between the patron and EZproxy to be encrypted, EZproxy requires the use of a large number of 
nonstandard ports or wildcard certs~\cite{le_wildcard}. We also believe that the absence of wildcard certificates can potentially 
lead to 2 major scalability issues in the Certificate Transparency logs. First, it can create an excessive storage overhead and also
as Certificate Transparency Logs are often distributed and replicated around the world, the lack of wildcard certificates can 
harm the performance of log synchronization when new certificates are added. 

\subsection{Geographical Analysis}

In order to answer the question of whether Let's Encrypt has facilitated the democratization
of TLS, we characterize the geographic distribution of Let's
Encrypt certificates. We focused on two 
key questions: (i) finding the countries where certificates issued by Let's Encrypt are most 
popular and (ii) finding those countries in which Let's Encrypt certificates have been disproportionately
popular compared to the overall number of websites (HTTPS enabled or not). 

\mypara{Analysis approach} 
To determine and visualize the geographical distribution of encrypted certificates
we use the following steps:

\begin{enumerate}
\item Collect Let's Encrypt domain names.
\item Resolve the domain names into valid IP addresses.
\item Determine Autonomous Systems for each of the demands, in order
to identify when many websites might be using the same cloud hosting
providers.
\item Look up each IP address and a geolocation database to get the latitude,
longitude, city, state and country.
\item Cluster data by country and visualize using a Choropleth map~\cite{choropleth}.
\end{enumerate}

As an aside, we observed that the process for resolving around 2 million domain names into IP addresses
pushes the limits of Domain name resolution as  DNS for end-users is a
relatively slow process that suffers in reliability when querying at high
speeds. To address this problem, we created a DNS query script written in
Go-lang that manually crafted DNS packets to hardcoded domain name servers.
Since each domain name server put query rate limiters on the program, the
system had to be designed to request DNS resolutions from many DNS servers at a
time, including multiple DNS resolvers from Amazon Web Services and Google. Ultimately, the
system was able to achieve over 500 DNS resolutions per second. This enabled us
to resolve 2 million domain names in one hour on a well-connected
server. 

 The method described above showed that 29,372 domains or 1.5\%
of the domains queried did not resolve to any IP addresses because the servers
had been taken down or are not publicly accessible. We consider 1.5\% of the
total number of domains to be a small fraction that does not skew the results
of the geographic distribution and exclude these from the analysis.  However, of the websites that did not resolve,
it was interesting to see how many of them had issued more than one Let's
Encrypt certificate. We found that 20\% of these domains had more than one
Let's Encrypt certificate.

\begin{figure}[h]
  \centering
    \includegraphics[width=200pt]{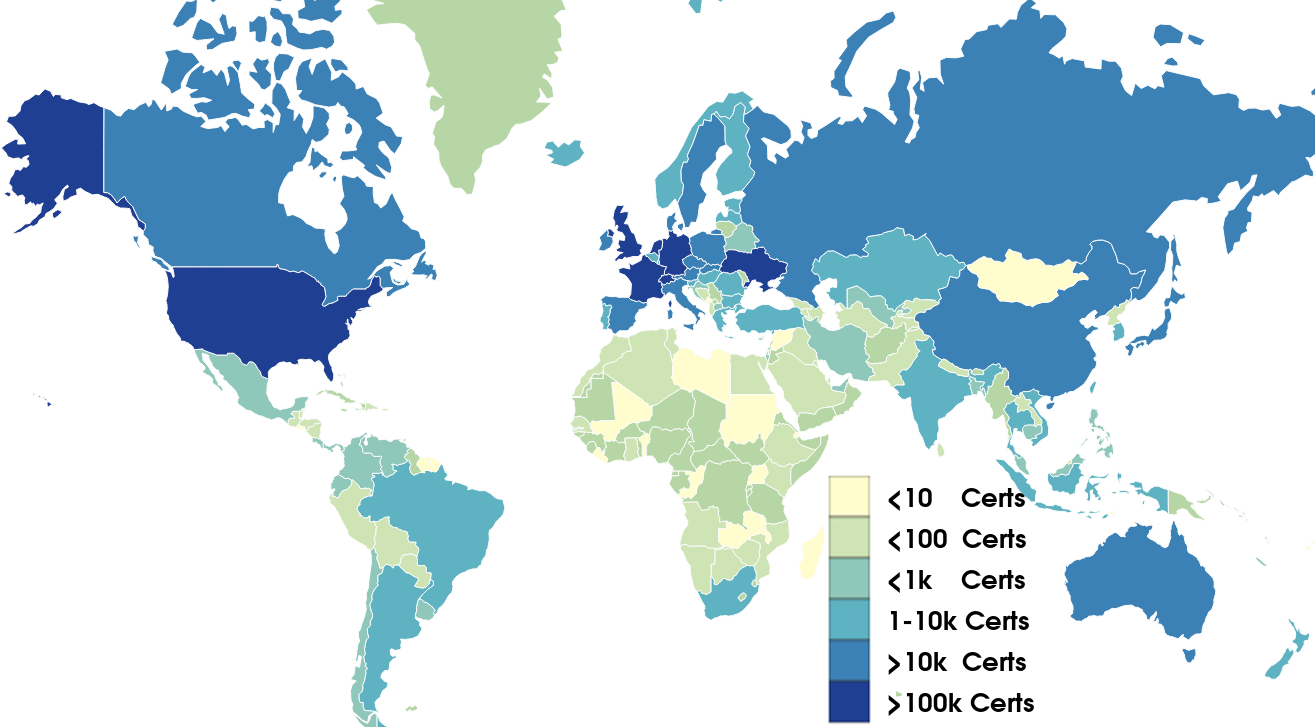}
    \caption{Raw Count of Let's Encrypt certificates}\label{fig:raw}
\end{figure}

\mypara{Results}
As Figure~\ref{fig:raw} shows, Let's Encrypt certificates are most 
commonly used in countries with high Internet penetration.
This is a seemingly obvious result but we delved deeper in order to find out
in which countries Let's Encrypt certificates are most popular. We compared the
number of active Let's Encrypt certificates with the active Internet population
of each country~\cite{internetstats}. The results can be seen in Table~\ref{table:cert}. 
Table~\ref{table:rawrank} shows the 20 countries with the highest number of issued certificates
as a percentage of the total number of Let's Encrypt certificates.
We observe that Let's Encrypt certificate penetration is predominant in Central 
and Western Europe and for countries like Switzerland, the Netherlands, France or
Germany the percentage of active usage and acquisition is 4 times higher compared to 
the United States. The ratio is 8 times for Canada. We postulate that
this fact is because a large number of Western European countries have
solid awareness about the importance 
of privacy on the Internet. Is a longstanding priority for the citizens of these countries 
and those considering hosting personal or business sites seek out
getting certificates from authorities like Let's Encrypt.    

\begin{table}[H]
\small
\begin{center}
 \begin{tabular}{||c | c||} 
 \hline
 Country & certificates$/1000$ users   \\ [0.5ex] 
 \hline\hline
 Switzerland & 5.0  \\ 
 \hline
 France & 3.887  \\
 \hline
 Netherlands & 3.703 \\
 \hline
 Germany & 3.061 \\
 \hline
 Singapore & 2.941  \\
 \hline
 Ireland & 2.338  \\
 \hline
 Czech Republic & 1.633 \\
 \hline
 Ukraine & 1.622 \\
 \hline
 Iceland & 1.489 \\
 \hline
 Luxemburg & 1.445 \\
 \hline
 Estonia & 1.172 \\
 \hline
 USA & 1.094 \\
 \hline
 Austria & 0.881  \\
 \hline
 Sweden & 0.828 \\
 \hline
 Saint Kitts and Nevis & 0.699 \\
 \hline
 Denmark & 0.635 \\
 \hline
 Norway & 0.535  \\
 \hline
 Canada & 0.525 \\
 \hline
 Slovenia & 0.523 \\
 \hline
 Latvia & 0.523 \\ [1ex]
 \hline
\end{tabular}
\caption {Top 20 countries with highest number of Let's Encrypt certificates 
as a function of the active Internet population}\label{table:cert}
\end{center}
\end{table}

\begin{table}[H]
\small
\begin{center}
 \begin{tabular}{||c | c||} 
 \hline
 Country & \% certificates/all LE certs   \\ [0.5ex] 
 \hline\hline
   United States & 28.14\% \\
 \hline
   Germany & 19.48\% \\
 \hline
   France & 19.41\% \\
 \hline
   Netherlands & 5.3\% \\
 \hline
   United Kingdom & 4.3\% \\
 \hline
   Switzerland & 3.27\% \\
 \hline
   Ukraine & 2.86\% \\
 \hline
   Slovak Republic & 2.22\% \\
 \hline
   Canada & 1.51\% \\
 \hline
   Japan & 1.39\% \\
 \hline 
   Czech Republic & 1.36\% \\
 \hline
   Singapore & 1.24 \% \\
 \hline
   Russia & 1.14\% \\
 \hline
   Ireland & 0.78\% \\
 \hline
   Sweden & 0.6\% \\
 \hline
   Australia & 0.63\% \\
 \hline
   Poland & 0.58\% \\
 \hline
   Austria & 0.55\% \\
 \hline
   Spain & 0.52\% \\
 \hline
   Italy & 0.33\% \\
 \hline
\end{tabular}
\caption {Top 20 countries with highest percentage of Let's Encrypt certificates 
as a function of total number of Let's Encrypt certificates}\label{table:rawrank}
\end{center}
\end{table}

Figure~\ref{fig:raw} also indicates that the number of Let's Encrypt certificates in 
smaller or less well connected countries is blooming and will likely continue into 
the future. Notable countries with surprising high rates of use are Ukraine, South Africa, 
Argentina and Brazil.

\begin{figure}[t]
  \centering
    \includegraphics[width=200pt]{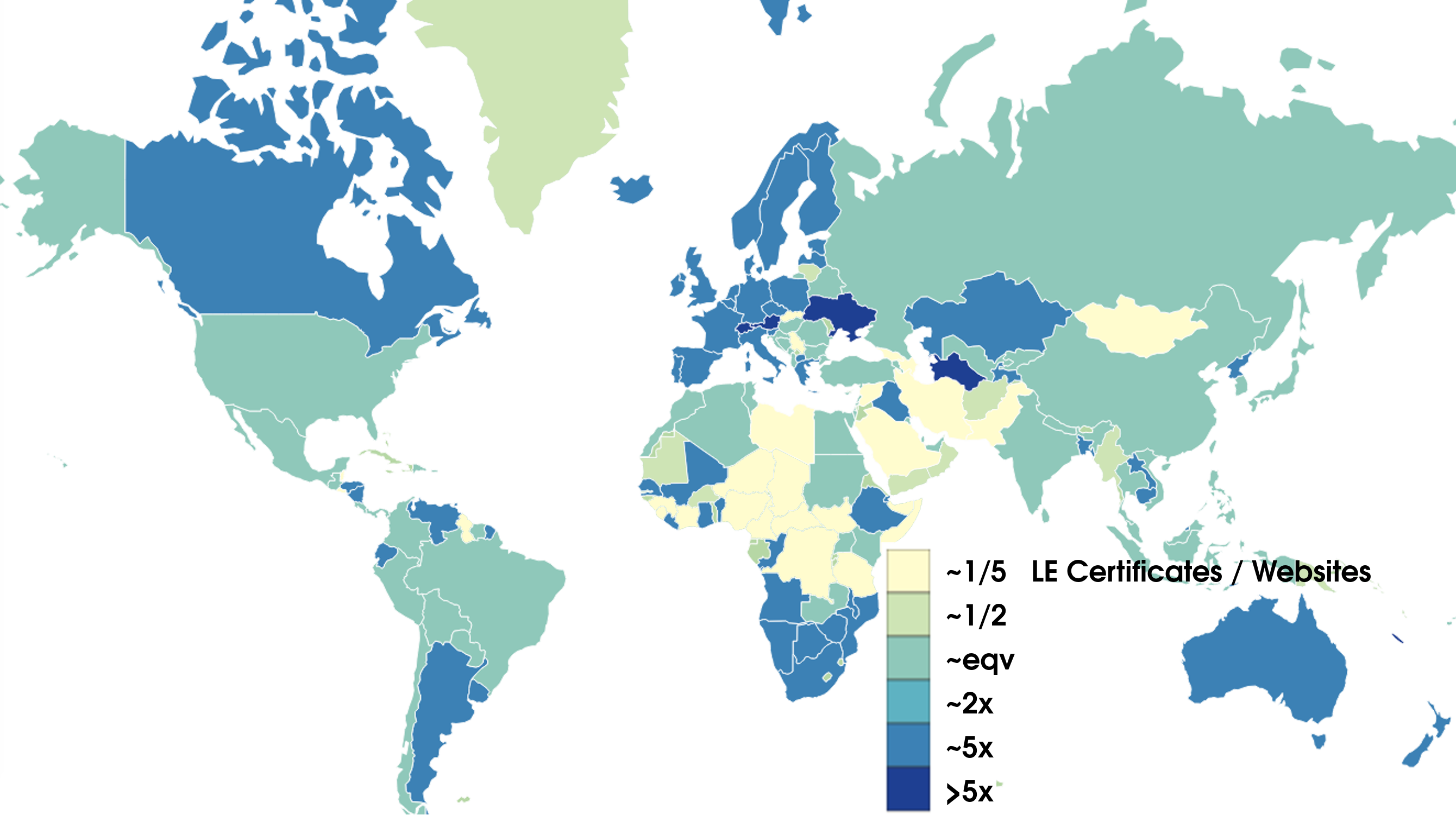}
    \caption{Relative popularity of Let's Encrypt certificates}\label{fig:map2}
\end{figure}

Another interesting question about geographic distribution is where Let's Encrypt
certificates are disproportionately popular compared to all website hosting or all certificates.
By comparing the number of websites in a particular country to the number of Let's Encrypt 
certificates in that country, we can determine the degree of market penetration of Let's Encrypt
certificates.   For each country, determine the number of any type of website and the number of Let's Encrypt websites hosted
there. We compute  the ratio of the number of Let's Encrypt sites compared to the number of sites
in that country, to indicate the relative popularity of Let's Encrypt there.
 For instance, If one country
had 1,000 sites in a sample of 1 million global sites, one would probably expect the number of Let's Encrypt certificates in that 
country to be 1,000 since they should be representing the same population (all websites). If, however, 
had over 5,000 certificates in the same sample size, one would be
surprised about how common Let's Encrypt certificates had been
deployed. This would be a positive sign for Let's Encrypt's value to people in those countries.
If, alternatively, the country had only a handful
of Let's Encrypt certificates, then it could raise the question of why are
users there not using Let's Encrypt as much as one would expect. This could be
due to language barriers on the Let's Encrypt site, appealing solutions offered
by traditional Certificate Authorities, or poor awareness and marketing of
Let's Encrypt in those countries.  Figure ~\ref{fig:map2} showcases the
aforementioned distribution of Let's Encrypt certificates worldwide. Notable countries
where Let's Encrypt is popular are Ukraine, Austria, Turkmenistan, Kazakhstan, Australia and South Africa.


%

\subsection{Domain History of using HTTPS}

Next, we  investigate the HTTPS characteristics of domains currently
using Let's Encrypt certificates. Specifically, we examine how many websites have used Let's
Encrypt as their first Certificate Authority and how many have transitioned from a traditional 
Certificate Authority to Let's Encrypt. To that end, we leverage one main source of data, Alexa 
records of the top 1 million domains. We acknowledge that this is a small subset of all domains 
however, given that many Certificate Authorities did not comply with Certificate Transparency until
very recently, the number of domains for which we could extrapolate information was extremely limited.
On the contrary, Alexa historic records contain information about the certificate chains of the most popular domains
and therefore they constitute a more credible source of data, even if
the sample of domains is smaller.

For every domain with Let's Encrypt certificate, we set out to identify the first date of issuance 
of the Let's Encrypt certificate and then compare with previous
records. Specifically we: (1) collect a list of all domains with certificates issued by Let's Encrypt;
 (2)  Find the earliest Let's Encrypt certificate issuance day for each of the domains, which can be found in the 
Certificate Transparency Log; and (3) Search the Certificate Transparency Logs for the aforementioned domains looking for certificates 
issued before Let's Encrypt certificates were adopted. 

We observe that in total, 503 domains transitioned from known CAs to Let's
Encrypt. Table~\ref{table:pop_cert} shows the top 5 Certificate Authorities
that clients left for Let's Encrypt.  An interesting insight is that most of
the CAs that users `abandoned' according to Alexa in order to transition to
Let's Encrypt, (not limited to CAs mentioned in the tables above) are CAs that
have one or more of the following characteristics: (1)  Have received poor
reviews from users in SSL reviews websites such as {\tt sslshopper}~\cite{sslshopper} and
{\tt spiceworks}~\cite{spicework};\ (2) Are among the Certificate Authorities that provide
`affordable' certificates~\cite{cheapcerts}; (3) Are not (or no longer)
trusted by popular browsers; or (4)  Have been involved in
security breaches~\cite{infoworld}.  This suggests  that Let's Encrypt has
become an attractive alternative for users who value the trust and privacy
provided by HTTPS but they are currently using untrusted CAs or are reluctant
to pay large amounts of money in order to obtain a certificate from one of the
more expensive Certificate Authorities.

\begin{table}[H]
\small
\begin{center}
 \begin{tabular}{||c | c||} 
 \hline
 CA & \% Transitioned to LE   \\ [0.5ex] 
 \hline\hline
 COMODO RSA CA & $31.21\%$ \\
 \hline
 StartCom CA & $12.3\%$  \\
 \hline
 RapidSSL SHA256 CA  & $9.7\%$  \\ 
 \hline
 Go Daddy  & $7.4\%$   \\
 \hline
 GlobalSign SHA256 CA & $2.7\%$   \\
 \hline
\end{tabular}
\caption {Certificate Authorities popular domains left to transition to Let's Encrypt}\label{table:pop_cert}
\end{center}
\end{table}

\subsection{ Adoption by popular domains}

\begin{figure}[h]
  \centering
    \includegraphics[width=200pt]{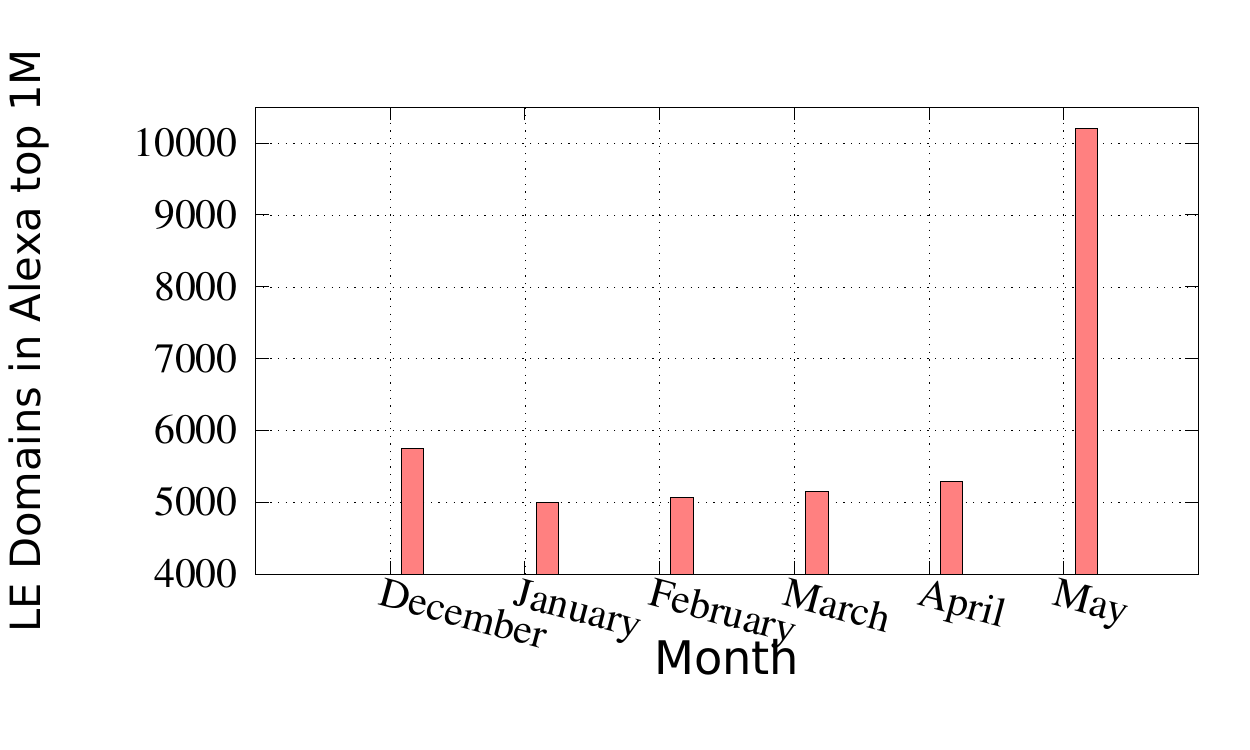}
    \caption{Number of Let's Encrypt certified domains in Alexa top 1M}\label{fig:alexa}
\end{figure}

For this analysis, we focus only on domains listed on Alexa top 1 million
domains and our goal is to investigate how popular Let's Encrypt is among them.
Figure~\ref{fig:alexa} shows how many Let's Encrypt websites are in Alexa's top
1M list of websites. We observe that even though there is a slight increase in
the total number of high profile websites that use Let's Encrypt, the rate of
acquisition is lower compared to the overall rate of acquisition of Let's Encrypt as
a CA.  Clustering those domains in groups based on their ranking in May showed
that 0.4\% are in positions between 1000 and 10,000. 7.7\% are in positions
between 10,001 and 100,000 and finally 92\% of these domains rank below
100,000.  The low rate of acquisition can be attributed to the fact that, a website with high
visibility on the web is most likely already certified by a CA; until its
current certificate expires, its administrators would not be particularly
motivated to switch from their current CA to Let's Encrypt.

\subsection{Summary of key observations}

To summarize, the key observations on the acquisition are :
\begin{enumerate}
\item 7\% of all Let's Encrypt certificates are requested by companies providing their clients with appliances like
like routers, modems, NAS or dynamic DNS servers. Companies like Synology, Free and DuckDNS take advantage of the free
automated Let's Encrypt certificates and acquire them in bulk in order to guarantee a secure channel between the client 
and the web interface of the client-owned device. 
\item More than 55\% of all Let's Encrypt certificates have been
  issued in Western Europe. Countries like Switzerland,
France and the Netherlands have an average of 3.9 certificates per 1000 Internet users. Furthermore, Let's Encrypt has 
democratized the procedure of obtaining TLS certificates worldwide and therefore in countries like Argentina or Ukraine 
Let's Encrypt certificates are over-represented in the population of HTTPS websites by a factor of 5.
\item Anecdotal evidence from Certificate Transparency logs suggests that Let's Encrypt has become an attractive alternative
for users who value the trust and privacy provided by HTTPS but are using either untrusted CAs or are reluctant to pay large 
amounts of money in order to obtain certificates from one of the more expensive Certificate Authorities. 
\end{enumerate}

\section{Analysis of Usage Characteristics}

In this section, we explore how domain owners are using certificates 
issued by Let's Encrypt ``in the wild''. We estimate the percentage of domains 
that have obtained multiple, redundant Let's Encrypt certificates as well as 
the percentage of domains that obtained but never renewed their Let's Encrypt 
certificates. We then run active tests in order to inspect how many 
domains actually use their Let's Encrypt certificates during the HTTPS handshake. 
Additionally, in an effort to detect malice in domains certified by Let's Encrypt, we leverage the 
VirusTotal service. Finally, investigate cases of typosquatting domains that are trying 
to exploit end-users' trust to HTTPS certified domains.

\subsection{Active usage analysis}

While Let's Encrypt's commitment to certificate transparency \cite{LE} provides
an unprecedented and valuable view into the TLS ecosystem, it can only serve as
an upper bound for expectations.  The mere issuance of a signed certificate
implies nothing about the certificate's deployment or utilization over time.
One data source that provides useful information about real world deployment is
the Censys.io search engine which is powered  by regular Internet-wide
scans~\cite{Durumeric}. Censys not only performed full IPV4 scans but also
conducts service discovery and follow on protocol handshakes to collect
amplifying information which can be used to validate the live deployment of
issued certificates.

To this end, we leveraged Censys in order to figure out the gap between issuance of Let's Encypt
certificates and their actual deployment. The resulting discrepancy between these
two values, encouraged us to investigate the case of
end users/early adopters that deploy Let's Encrypt certificates out of curiosity.
We measure cases of duplicate certificates issued, and we also measure the percentage
of certificates that were issued and never renewed. This information comes exclusively
from Certificate Transparency Logs and Alexa records. As a final step, we perform 
our own active testing on the domains that claim to be using a certificate issued by Let's 
Encrypt in order to actually verify our results.

\begin{figure}[H]
  \centering
    \includegraphics[width=250pt]{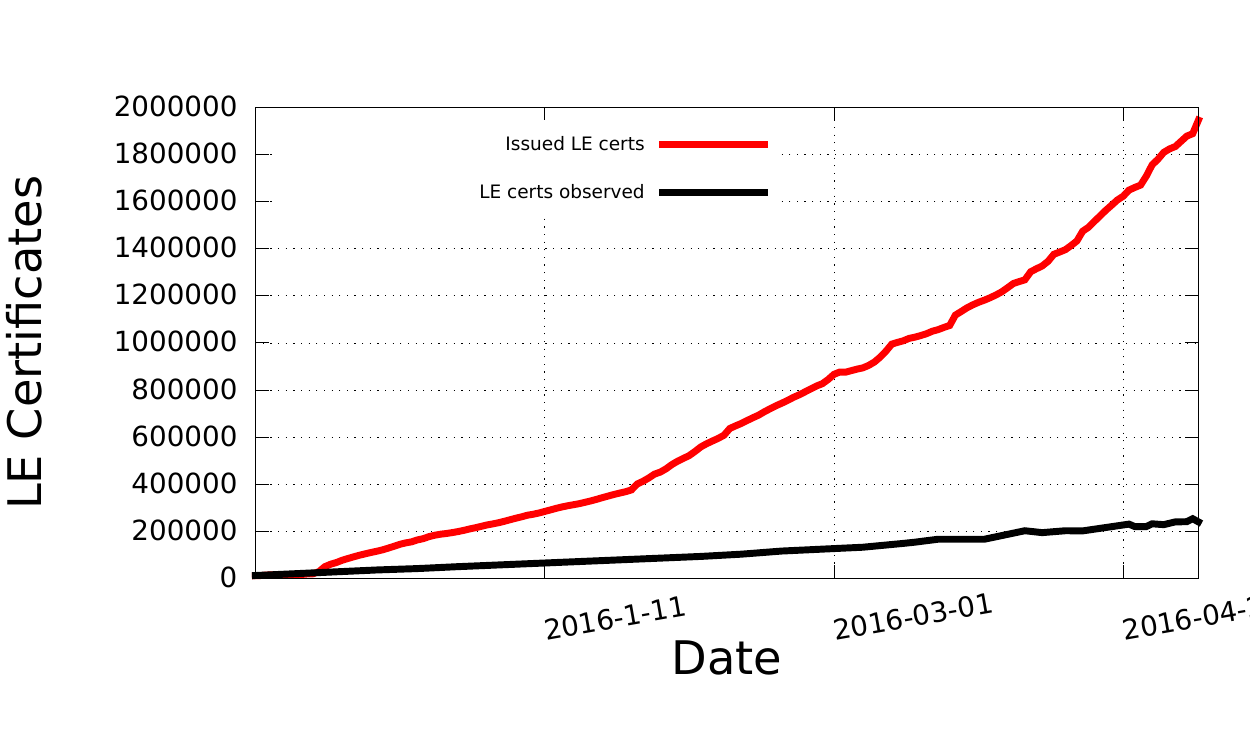}
    \caption{Let's Encrypt certificates observed by Censys}\label{fig:censys}
\end{figure}

Our results, shown in Figure~\ref{fig:censys}, show that the highest observed total number of 
certificates does not exceed 253,892 on May 3rd 2016. This is an order of magnitude 
lower than the naive upper bound discerned from Certificate Transparency logs of 
2,799,771 and the more realistic bound of 1,932,242 unexpired certificates as 
reported by Let's Encrypt. 

Now, some of these may reflect potential gaps in Censys scans. To this end, we
also conducted our own active HTTPS scanning.  By actually attempting to
resolve a domain from the Certificate Transparency Logs and making an active
HTTPS request we can safely assert which certificates are active. This
addresses the overcounting problem of Certificate Transparency logs and
potential undercounting by Censys. We collected all the
unique names from the transparency logs and then ran the {\tt sslyze}~\cite{sslyze} tool,  a full featured SSL scanner
written in Python, against each of them.\footnote{One caveat is  that sslyze does not
take into account SAN domains~\cite{san} and only attempts to connect to port 443.}

\begin{figure}[H]
  \centering
    \includegraphics[width=250pt]{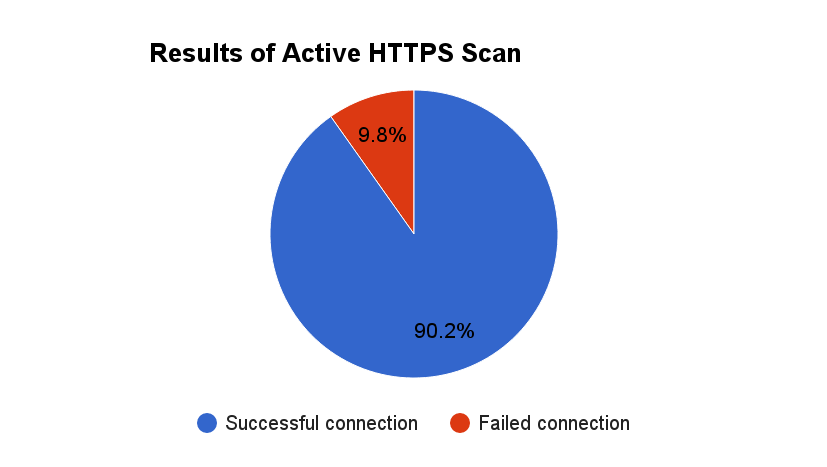}
    \caption{Proportion of sites that successfully initiated HTTPS connections}\label{fig:success}
\end{figure}

The results of this test are summarized in
Figures~\ref{fig:success}---\ref{fig:active}. We  observe that 90.2\% of our
domains successfully completed a HTTPS handshake, whereas 9.8\% failed. If we
take a closer look at the reasons behind the failures we will observe that
16.8\% of them can be attributed to DNS failures whereas the majority of failures 
is due to timeout errors, rejected or incomplete handshakes.\footnote{The reason behind the high number of timeouts is that
SSLyzer only attempts connections to port 443 and does not follow URL redirects.
That adds some inaccuracy to the results because as we observed, when we tried
to access freeboxos domains, they generated a 302 HTTPS redirect message and
thus we categorized them as failures.}

 The most interesting observation however
is that of the active domains owing a certificate by Let's Encrypt, only 54\%
are using that certificate during the TLS handshake. We delved deeper in order
to inspect what domains were serving different certificates and what
Certificate Authority those certificates were issued from.
Table~\ref{table:diffca} shows the results for the top 20 eTLDs that had acquired 
 but were not using Let's Encrypt  certificates.

\begin{figure}[H]
  \centering
    \includegraphics[width=250pt]{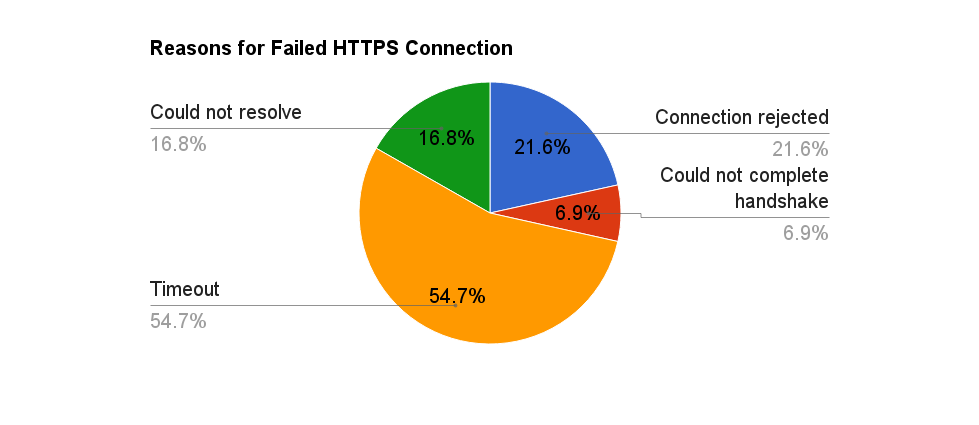}
    \caption{Reasons for failed HTTPS connections}\label{fig:failed}
\end{figure}

\begin{figure}[H]
  \centering
    \includegraphics[width=250pt]{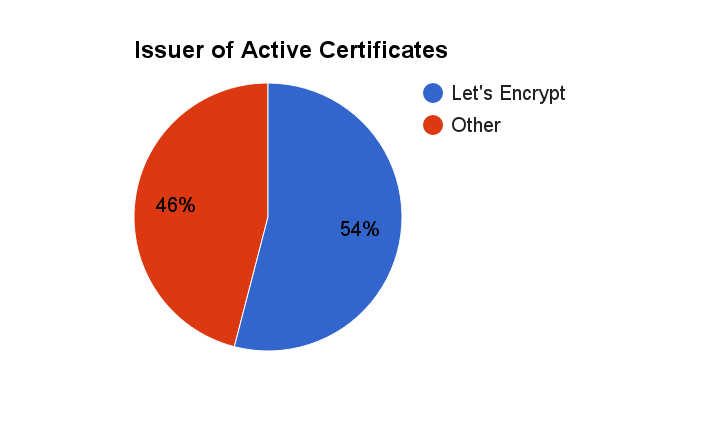}
    \caption{Proportion of sites with active Let's Encrypt certificates as a proportion of all sites that own a Let's Encrypt certificate}\label{fig:active}
\end{figure}

\begin{table}[h]
\small
\begin{center}
 \begin{tabular}{||c | c | c||} 
 \hline
 eTLD & CA & \#domains  \\ [0.5ex] 
 \hline\hline
 ovh.net &AlphaSSL-SHA256  & $359757$  \\
 \hline
 ovh.net &AlphaSSL& $307565$  \\
 \hline
 cloudflaressl.com & COMODO ECC & $49932$  \\ 
 \hline
 elbiahosting.sk & *elbiahosting.sk & $8549$   \\
 \hline
 hosting-admin.net & ssl.hosting-admin.net & $4503$   \\
 \hline
 synology.com & Synology Inc. & $1354$ \\
 \hline
 dreamhost.com & sni.dremhost.com & $1287$ \\
 \hline
 kasserver.com & COMODO RSA & $797$ \\
 \hline
 planroomcheckout.com& COMODO RSA & $711$ \\ 
 \hline
 pagekite.me & StartCom Class 2 & $406$ \\ [0.5ex]
 \hline
\end{tabular}
\caption {top 10 eTLDs and number of domains owning Let's Encrypt Certificates but not serving them}\label{table:diffca}
\end{center}
\end{table}

Interestingly, we find 
 that most of these are  hosting services which were acquiring a large number of certificates 
 figures prominently in this list. 
The most interesting example is again the one of {\tt ovh}, a France based cloud and web hosting provider, also a platinum sponsor 
of Let's Encrypt~\cite{platinum}. {\tt ovh} promises to provide standard encryption for his services and to use Let's Encrypt to that
end. The fact that they are using AlphaSSL shows that either they have misconfigured their deployment (a rather probable
result after manual inspection of selected {\tt ovh} hosted websites) or they are still in the process of transitioning from AlphaSSL
to Let's Encrypt. Another interesting observation is that many of them were using self-signed certificates; 
 e.g.,  {\tt elbiahosting} returns self-signed certificates even though on their website they claim that they have
started supporting Let's Encrypt certificates.  At this time, we cannot speculate 
 on their motives for not using the acquired Let's Encrypt certificates despite their publicly stated 
 intent to do so.

\subsection{Miscellaneous Characteristics of User Behavior}
Next, we shed light on three categories of anecdotal but non-trivial characteristics of user behavior
 that we observed during our  analysis.

\mypara{Redundant certificates} To originally explain the discrepancy between
the Censys results and the number of certificates, we posited that many domains
might be issuing redundant certificates.  This led us to examine the amount of
redundancy. The results showed that on average, 9\% of the exact domains in
the Transparency Log had 
issued more than one certificate at the same day, with that percentage reaching
around 30\% during the months of December 2015 and January 2016 when Let's
Encrypt was still at its first days as a stable product. This shows that to some extent the free nature
of Let's Encrypt certificates triggers the curiosity of users who experiment
with the platform and issue more certificates than they need. Even though this
has no implications to the actual website as it is up to the webmaster to
choose which certificate will be served, a potential implication of that is the
unnecessary bloating of Certificate Transparency Logs that now have to keep
track of unused certificates.

\mypara{Non renewal of certificates} To explore further the `curiosity' of end
users and their willingness to adopt certificates issued by Let's Encrypt we
also calculated the number of certificates that were issued but never renewed
(even though Let's Encrypt has a policy for automatic certificate renewal every
90 days). We observed that around 15\% of all issued certificates
were never renewed. We believe that this number describes potential adopters
who were willing to experiment with Let's Encrypt but decided not to continue
using its services.

\mypara{Toy/fake CA} A low-scale but still interesting observation about
potential misconfiguration of Let's Encrypt certificates came from a search in
the list of Certificate Authorities that popular domains had obtain
Certificates from. One of the CAs that appeared in that list was {\tt Happy 
Hacker Fake CA}. We investigated that CA further and discovered that it is
managed by the staging server of Let's Encrypt. This means that some end users
during their experimentation with Let's Encrypt successfully managed to obtain a
certificate, only to realize later (if at all) that this certificate was issued
by a fake CA. Even though browsers would mark this certificate as untrusted, we
believe that for an inexperienced user it would take time until he realized
that his website was insecure and untrusted despite his belief about the
opposite.


\subsection{Potential use for Malice}

Traditionally end users have been `trained' to trust a website if it has the verified certificate
symbol against its domain name. However, now with the existence of Let's Encrypt,
an adversary can easily obtain a valid certificate for a malicious or typosquatting
domain making an end user fall prey to phishing and drive by download category of
attacks.

\subsubsection{Hosting malware} We make use of the VirusTotal API to detect malicious 
activity on Let's Encrypt-signed websites. The VirusTotal API provides
a detection ratio depending on how many of these services classified a
particular domain as malicious. We were able toobtain a private API
key  to conduct this measurement and made use of the URL report API
to scan a sample of 100,000 domains from the unique domains in our
Certificate Transparency database. The reason why we only chose a
small  (but random) sample of domains was our limited
quotas on VirusTotal along with the high latency of executing URL scans. Virus Total 
allowed us to inspect domain names using 67 different antivirus solutions and reported 
how many out of the 67 solutions classify the website as malicious.

An interesting first result of our VirusTotal measurements is that 82,524 of the 
100,000 sampled domains were scanned for the very first time. This implies that end users, web 
administrators and even certificate transparency monitors fail to run safety checks on
websites to ensure protection from malicious actors. 
A second scan was therefore conducted to gather the results for all the domains that were scanned
for the first time. 

\begin{figure}[h]
  \centering
    \includegraphics[width=200pt]{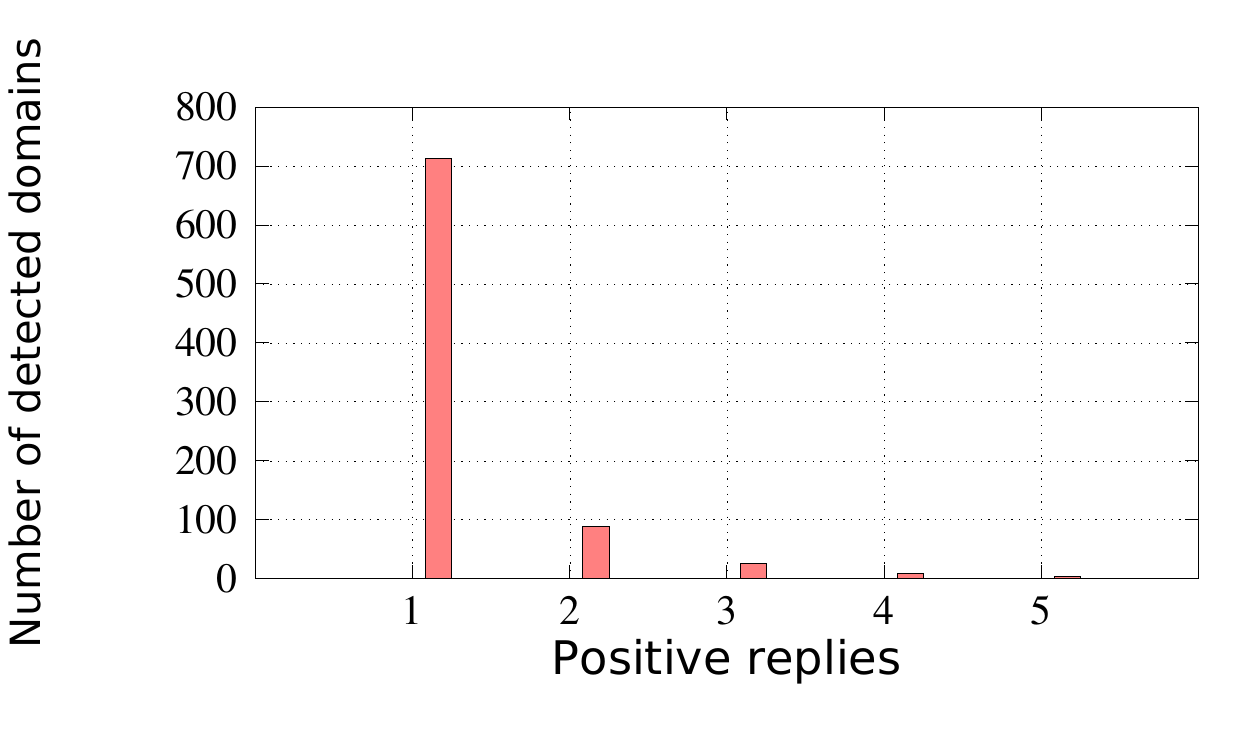}
    \caption{Number of websites with Let's Encrypt certificates that triggered one or more antivirus alerts through Virus Total}\label{fig:vt}
\end{figure}

Figure~\ref{fig:vt}  shows that around 1\% of the websites inspected were found to be malicious 
by one or more antivirus solutions. Although the total number of domains identified is a very 
small component of the total number of domains scanned (and of the broader HTTPS ecosystem), we believe 
that this finding constitutes an indication that malicious adversaries have started to take the ease of 
issuing TLS certificates as an opportunity to exploit end users' notion of trust.

\subsubsection{Typosquatting}

As a next step we focused on domains that were most likely being typosquatting~\cite{Khan,KULeuven}. 
To determine which of the 
Let's Encrypt sites are likely typosquatting, we leveraged 
 a technique based on the Levenshtein distance\cite{levenshtein}, which measures   how similar 
two strings of characters are, where the measure is defined in terms of an 
`edit distance'. Edit distance describes the number of modifications to one 
string that you need to get the other. By using this technique in a 
generative fashion, we can create sets of domain names that are 
a low edit distance away from popular brand names. We searched for domains 
with close matches (strings with an edit distance of maximum one).

 Specifically, we take  the Alexa top 250 domain names. We remove those that are
less than 6 characters or too common (for example `apple' as the word is too common). Then,
we ran a Levenshtein distance generator with a maximum edit distance of one and
removed anything that ended up as a real word (e.g. flickr -$>$ flicker).
Thus, we determined a list of domain name misspellings that are similar to
other common domains. Then, performed a lookup of this large list of possible
misspellings to determine which of them actually appear in the set of Let's
Encrypt signed certificates. We ignored changes in the told level
domain.  We were careful to remove sites that were
obviously not typosquatting, in case they were included in the list due to
spelling similarities.

\begin{figure}[H]
  \centering
    \includegraphics[width=200pt]{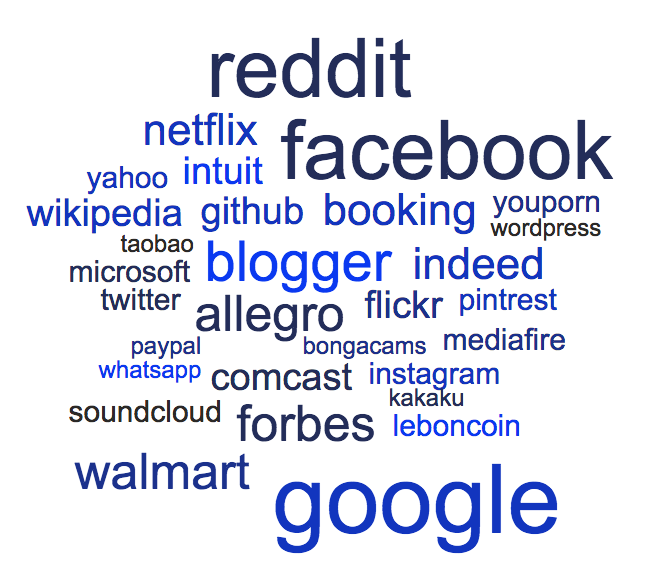}
    \caption{Top sites targeted by typosquatting within Let's Encrypt
      Certificate population}\label{fig:active}
\end{figure}

Out of the 105 eligible popular domain names sampled, we discovered
203 likely typosquatting domain matches (e.g. domain within an edit
distance of one from the original).  We also had to be careful to remove sites that were
obviously not typosquatted, in case they were included in the list due to
spelling similarities. A sample of the most commonly targeted domains are
shown in Table~\ref{table:close} along with example URLs.

\begin{table}[h]
\small
\begin{center}
 \begin{tabular}{||c | c | c||} 
 \hline
 Domain Name & Number matches  & Example Domain   \\ [0.5ex] 
 \hline\hline
 Google & $14$ & googlez.fr \\
 \hline
 Facebook & $12$ &  faecbook.fr \\
 \hline
 Reddit  & $12$ & redditq.com  \\ 
 \hline
 Blogger & $8$  & blogges.de  \\
 \hline
 Forbes & $7$ & dorbes.com   \\
 \hline
 Allegro & $7$ & alleguo.com \\
 \hline
 Netflix & $6$ & netflix.desi \\
 \hline
 Booking & $6$ & bookhing.com \\
 \hline
\end{tabular}
\caption {Sample of top typosquatted target and example typosquatting domains}\label{table:close}
\end{center}
\end{table}

\subsection{Observations}

To summarize, the key observations are :
\begin{enumerate}
\item We investigated the actual usage patterns of Let's Encrypt certificates in order to assess how users 
interact with the service. We observed that on average 9\% of Let's Encrypt domains have obtained multiple, 
redundant certificates and 15\% of all the certificates obtained have not been renewed. 
\item We ran active HTTPS tests to evaluate how many of the domains with an unexpired certificate are actually
serving their clients with that certificate. Surprisingly enough, only 54\% of the active domains use their 
Let's Encrypt certificate whereas 46\% use either self-signed certificates or certificates issued by other 
CAs. 
\item To detect malice in the domains we leveraged the VirusTotal service which examines each domain against
67 antivirus solutions and found that 1\% of the domains, tested positive for some threat.
\item Further analysis showed that malicious actors are attempting to exploit end users' trust by using 
certificates issued by Let's Encrypt for domains that are either typosquatting.
\end{enumerate}

\section{Implications}

In this section, we discuss the potential  implications of our 
 observations for different players in the HTTPS ecosystem: 
 Let's Encrypt,  web domains,  and end-users and  browsers.

\mypara{Let's Encrypt} We identify a few key opportunities for Let's Encrypt 
 to improve adoption  and also improve overall efficiency and  scalability. 
 Specifically, 
\begin{itemize}
\item {\em Revisit the need for wildcards:}  With more and more
companies issuing individual Let's Encrypt certificates for their clients, and
with more and more users  issuing multiple certificates out of
curiosity or need,
Let's Encrypt should consider the option of allowing its users to choose whether they want to issue
 wildcard certificates or not. It could  potentially enforce quotas for individual domains. 
  The benefit of these
actions is clear: Certificate Transparency logs will remain manageable in size
as with the current trends, adding around a million certificates on a monthly
basis can potentially lead to scalability issues for these logs and performance
issues for monitors that want to inspect them or a regular basis. 

\item {\em Check intent before issuing certificates:}  Being free 
 and automated lowers the barrier not only for less popular websites
 but also for abusive purposes. We already see early evidence of the use 
 of Let's Encrypt for  malware-laden websites and typosquatting. One option 
 is that Let's Encrypt can use well-known tools and approaches from the security  
 community to check if the website has potentially abusive 
 intent before issuing a new certificate; e.g., look at the Google SafeBrowsing history 
 of the site or initiate an active scan on a new certificate request.      While this 
 does not offer perfect protection against users (e.g., a newly created domain  can be dormat and turn malicious later), 
 this does raise the bar for websites that are already known to exhibit malicious intents. 

\end{itemize}

\mypara{For Users and Client Browsers} In terms of users and user-facing 
 browsers, we identify two key implications:
\begin{itemize}
\item {\em Protect the User through Website Analysis}  The increasing adoption
of HTTPS definitely leads to a safer and more trusted browsing environment for users who share 
more and more of their private data online. However, the ease of deployment of Let's Encrypt
certificates has shown that it becomes easier for an attacker to exploit that
notion of trust in order to deploy `trusted' malicious domains. Specifically, Let's Encrypt as it 
is lacks supervision in the domains that it signs; it can contribute to the creation of seemingly 
 secure  malware-hosting and typosquatted domains.  For instance, given that we observe many 
 popular websites are not using Let's Encrypt, it might be easy to distinguish typosquatting behaviors 
 against popular domains by simply looking at the certificates. 

\item {\em Proactively check for configuration errors:} Our
  observations suggest that there are 
 serious misconfigurations among many website owners who use 
 Let's Encrypt. Additional documentation, training materials, and native
 language support could educate these users on proper
 deployments. Additionally, scanning sites for misconfigurations and
 alerting the user could provide valuable encouragement for site
 owners to improve.
 From the users side, additional transparency about deployment quality
 can allow users and browsers to proactively check for dangerous
 connections when accessing newly HTTPS-enabled sites. 

\item {\em Use active measurements to complement Certificate Transparency:}  
 We observe that nearly half of all issued certificates are never used, including by 
 domains that are purportedly sponsors of the the Let's Encrypt effort. This suggest that
 looking at the transparency log by itself may not be a sufficient proof of the ``active'' 
 status of a certificate and thus we will still need some active measurement 
 system~\cite{bates-imc14,wendlandt-usenix08} to complement transparency logs.

\end{itemize}

\section{Conclusions}
 
The emergence of Let's Encrypt and Certificate Transparency are promising and
potentially revolutionary trends in the HTTPS ecosystem.  To fully 
unleash the potential of these opportunities, however, we need to 
 have a systematic understanding of how users are {\em acquiring}  and 
 actually {\em using} these capabilities. This measurement study is 
 a first look at the adoption characteristics of
 this emerging ecosystem.  

In particular, we find  that  Let's Encrypt is being acquired more by
less-popular domains and in countries with traditionally lower Internet
penetration, which suggests the potential to democratize the benefits of HTTPS.
At the same time, however, we do observe a certain degree of ``lack of
seriousness'' in adoption; many certificates are inactive and
there are obvious sources of misconfigurations and inefficiencies in
how deployed certificates
are being used.  Finally, we also make a cautionary note that ease of acquisition and low cost
is a double-edged sword. As it improves accessibility, it also lowers the barriers for malicious uses.
We see early evidence of  potential sources of abuse of Let's Encrypt by 
 malicious websites hosting malware and/or typosquatting. 

We discussed key implications for different players in this ecosystem and how 
 they can use these findings to take actionable measures. For instance, Let's Encrypt 
 could offer new capabilities to  offer wildcard   certificates to
 reduce load. There may also be a need to  
 check for abusive intent through 
 a community-run warning system. 
Finally, browser vendors and users 
 can take easy precautions to detect misconfigured deployments to avoid privacy violations.   

As with any study looking at an emerging phenomenon, 
 our datasets invariably have some limitations and biases. Despite these, 
 we believe that our analysis has shed light on some key observations 
 on the adoption. These have important implications that
 can inform the future of these technologies. We hope that our analysis 
 and results inspire future measurements and can translate into action items
 for various parties involved. 

\comment
{Let's Encrypt is a certificate authority that is encouraging the average user
to adopt website encryption for the first time.  This has already proven to
increase web security on the internet and the pace does not appear to be slowing.
Over two million certificates have now been signed by Let's Encrypt, and for
the first time, the community has a CA with an almost complete, public list of all of these certificates
published by the CA itself. Let's Encrypt's free certificates are one
innovation in the security space, and are a boon for price-conscious users.
Their automated, fast signing process is lowering the barriers to entry for
many first time HTTPS users. The most important takeaway from Let's Encrypt's
example is, however, the concept of Certificate Transparency. By publishing a
complete list of all the certificates that have been signed by the authority,
the entire world can monitor for malicious domains, typo-squatted domains,
incorrect server configurations, and more, all without relying on trusted
certificate authorities to do this behind closed doors. And as Facebook found
recently, exposing the complete set of certificates has practical purpose in
finding fraudulent, typosquatted domains sending phishing emails. As Let's
Encrypt and the Certificate Transparency movement grow, may the industry
continue to say, `Let's Encrypt All the Things'.
}

{
\small
\bibliographystyle{abbrv}
\bibliography{vyas}
}

\end{document}